# Peripheral Nitrogen Topology as a Defect-Chemical Switch for Electronic and Magnetic States in Graphene: A First-Principles Study of Pyridinic, Pyridazinic, Pyrrolic, and Pyrazolic Configurations

Md. Moktadir Billah Tahmid[1], Indranil Rudra[2], Jahid Emon[3*], Mohammad Jane Alam Khan[1]

[1]Department of Mechanical Engineering, Bangladesh University of Engineering and Technology, Dhaka, Bangladesh

[2]Department of Mechanical and Aerospace Engineering, University of Alabama in Huntsville, Alabama, United States

[3]Department of Mechanical Science & Engineering, University of Illinois Urbana-Champaign, Urbana, IL, United States

[*]Corresponding authors: jemon2@illinois.edu

## Abstract

Defect and heteroatom engineering offer powerful routes for tuning the electronic and magnetic properties of graphene, yet the role of specific peripheral nitrogen topologies around graphene voids remains insufficiently understood. Here, spin-polarized first-principles calculations were performed to investigate how four heterocyclic-like peripheral nitrogen configurations- pyridinic, pyridazinic, pyrrolic, and pyrazolic modify the structural stability, charge redistribution, electronic structure, and magnetic response of graphene containing a central void. Among the four peripheral N configurations, the pyridinic N provides the most favorable structural-energetic balance among the investigated motifs. Bond-length analysis reveals that nitrogen topology strongly controls local lattice reconstruction. Charge-density, charge-density-difference, and Bader analyses demonstrate that the peripheral N atoms act as electron-accumulating centers and reshape the local electronic environment around the vacancy rim. Spin-resolved band structures show that pyridinic, pyridazinic, and pyrrolic configurations retain metallic or near-metallic defect-state character, whereas pyrazolic graphene opens a narrow band gap. Magnetic analysis further reveals that pyrazolic graphene is spin-compensated, with zero net magnetization, unlike the other systems, which possess finite spin-polarized moments. Spin-density and SPDOS analyses indicate that the magnetism originates from N-modulated vacancy-edge states involving both N 2p and neighboring C 2p orbitals. These findings establish peripheral nitrogen topology not merely as a structural defect descriptor, but as a deterministic defect-chemical switch, offering a metal-free route to pattern active spintronic and semiconducting domains directly into the graphene lattice through controlled vacancy-edge nitrogen coordination.

**Keywords:** Graphene; Nitrogen doping; Peripheral nitrogen defects; DFT; Band-gap engineering; Defect-induced magnetism

## Introduction

Graphene, a two-dimensional honeycomb lattice of $sp^2$-bonded carbon atoms [1], has attracted sustained attention because of its exceptional electronic structure [2], high carrier mobility [3], mechanical robustness [4], and atomically thin geometry [1]. Its low-energy charge carriers behave as massless Dirac fermions near the K and K' points of the Brillouin zone [2, 5], giving pristine graphene an unusual semimetallic character with a zero band gap [2, 5]. This electronic uniqueness makes graphene scientifically important, but the absence of an intrinsic band gap [6] and the lack of intrinsic long-range magnetism [7] limit its direct use in semiconductor logic, spintronic components, and magnetically active carbon-based devices. Therefore, controlled modification of graphene's electronic [6] and magnetic structure [7, 8] remains a central challenge in graphene materials science.

Defect engineering [9, 10], edge engineering [11], and heteroatom doping [12, 13] have emerged as effective routes for modifying graphene beyond its pristine semimetallic state. In particular, vacancy defects [10] and reconstructed graphene edges [11] can introduce localized electronic states near the Fermi level, and these localized states may generate spin polarization when sublattice imbalance [14, 15], dangling-bond states [9], or defect-induced $p_z$ -orbital localization occur [7, 15]. Such defect-induced magnetism is especially attractive because it suggests the possibility of producing magnetic behavior in a metal-free carbon framework [7], avoiding transition-metal impurities that may complicate stability, toxicity, or device reproducibility [8]. However, the emergence of magnetism in defective graphene is highly sensitive to local atomic geometry [9], defect reconstruction [10], charge redistribution [16], and orbital occupation near the Fermi level [15].

Nitrogen is one of the most studied heteroatom dopants [12] due to its resemblance to carbon in terms of atomic size [17] but has one valence electron more and a greater electronegativity [18]. Graphene incorporation of nitrogen can be altered by a variety of chemically distinct bonding conditions which often include- graphitic/substitutional N [13, 19], pyridinic N [13], pyrrolic N [19] and other defect-related nitrogen species. These are not electronically equivalent nitrogen structures [13], graphitic nitrogen typically replaces a carbon atom in the basal plane, and pyridinic and pyrrolic nitrogen may be found at vacancy edges, defects, or non-hexagonal local bonding structures [12]. The structural identity of nitrogen can be more significant than the overall nitrogen content because each type of nitrogen has a differing effect on local bonding [13], charge density [16, 20], and orbital hybridization [19].

The first-principles studies have revealed that the formation energy [21], structural stability [22], electronic structure [13], and magnetic response [22, 23] of N-doped graphene are highly sensitive to the dopant structure and defect topology. On the same note, in-plane defects of graphene have been studied using DFT, and it has been shown that the nitrogen structures of vacancies can significantly modify stability [9], electronic states [15], and magnetic moments [7, 10]. The magnetic properties of N-doped graphene are particularly important for carbon-based spintronic design because nitrogen can tune the spin-dependent electronic structure [22, 23] without the addition of transition metals [8]. It has been reported that N-doped graphene exhibits ferromagnetic behavior at a high Curie temperature [24], and that nitrogen incorporation can produce technologically-relevant magnetic responses in graphene under appropriate structural conditions [24].

Although much progress has been made, the majority of computational investigations have predominantly involved graphitic, pyridinic, and pyrrolic nitrogen structures [8], but little has been done to address more specific peripheral nitrogen structures that are reminiscent of organic heterocyclic structures around a graphene defect. This gap is important because nitrogen atoms located at a vacancy perimeter may form local bonding patterns analogous to azine- and azole-type heterocycles, and these motifs can impose different degrees of C-C distortion, C-N bond asymmetry, charge compensation, and spin polarization. Pyrazolic-structure-rich nitrogen-doped graphene has also been experimentally explored for energy-related applications [25], showing that pyrazolic nitrogen environments are chemically meaningful and not merely theoretical constructs. The formation of pyridazinic N at the edge of the void is also identified [26], highlighting its structural preference for defective regions. However, a systematic first-principles comparison of pyridinic, pyridazinic, pyrrolic, and pyrazolic peripheral N motifs around a graphene void is still limited. In particular, the link between local heterocyclic-like N topology and structural reconstruction, charge transfer, band-structure modulation, orbital-projected density of states, and magnetic moment formation remains insufficiently clarified.

Here, we present a spin-polarized first-principles investigation of four peripheral nitrogen configurations in graphene: pyridinic, pyridazinic, pyrrolic, and pyrazolic N-doped vacancy structures. Unlike conventional substitutional N-doped graphene, the systems studied here contain nitrogen atoms located at the boundary of a central graphene void, allowing the role of peripheral N topology to be examined directly. The computational protocol is first validated using a standard substitutional nitrogen-doped graphene model, after which the four peripheral configurations are compared in terms of optimized geometry, C-C and C-N bond-length distortion, defect formation energy, charge-density distribution, charge-density difference, Bader charge transfer, spin-resolved band structure, spin-polarized projected density of states, N-orbital-resolved density of states, total magnetization, spin density, and local magnetic moments etc. The main objective of this work is to elucidate the effect of various peripheral nitrogen motifs on the structural, electronic and magnetic properties of defective graphene. The results show that the topology of the N topology around the void is a determining control parameter. These results provide a structure-charge-spin relationship linking the local arrangement of the peripheral nitrogen atoms to charge redistribution, defect-state occupation, orbital hybridization, and net magnetic response.

## Methodology

### *Construction of pristine and peripheral N-doped graphene models*

First-principles periodic models were constructed to investigate the influence of peripheral nitrogen coordination on the structural, electronic, and magnetic properties of nitrogen-doped graphene. The pristine graphene reference model was built using a 5 × 5 supercell containing 50 carbon atoms. The 5 × 5 supercell was selected to provide a practical balance between sufficient isolation of the localized vacancy-edge defect states and computational feasibility for high-precision spin-polarized band-structure, density-of-states, etc. calculations. A vacuum spacing of 12 Å was introduced along the z direction to minimize artificial interactions between periodically repeated graphene layers. Four peripheral nitrogen configurations were then generated from the same parent graphene supercell: pyridinic, pyridazinic, pyrrolic, and pyrazolic N-doped graphene. In each case, a central void was created by removing selected carbon atoms from the graphene lattice, followed by the incorporation of nitrogen atoms at the exposed defect boundary. For all configurations, the same supercell framework, computational parameters, and post-processing procedures were used to ensure a consistent comparison among the four nitrogen-defect motifs.

**Table 1:** Structural composition and defect stoichiometry of the configurations

| Configuration | C atoms removed | N atoms added | Final formula |
|---|---|---|---|
| Pyridinic | 6 | 2 | $C_{44}N_2$ |
| Pyridazinic | 7 | 2 | $C_{43}N_2$ |
| Pyrrolic | 7 | 2 | $C_{43}N_2$ |
| Pyrazolic | 6 | 2 | $C_{44}N_2$ |

***Density functional theory setup***

All calculations were performed within the framework of density functional theory [27] using the plane-wave pseudopotential approach [28] as implemented in the Quantum ESPRESSO package [29, 30]. The exchange-correlation interaction was described using the generalized gradient approximation with the Perdew-Burke-Ernzerhof functional [31]. The core-valence electron interactions were treated using Projector Augmented Wave pseudopotentials [32] obtained from the PSlibrary [33]. The kinetic energy cutoffs for the plane-wave basis set and charge density were set to 55 Ry and 440 Ry, respectively. Brillouin zone sampling was performed using a Monkhorst-Pack [34] $7 \times 7 \times 1$ k-point mesh for geometry optimization and self-consistent field calculations. A Gaussian smearing scheme with a broadening width of 0.01 Ry was used to improve convergence for metallic and near metallic spin polarized states. The self-consistent field convergence threshold was set to $10^{-6}$ Ry. All defective systems were treated using spin-polarized calculations in order to capture possible defect-induced magnetic states.

Geometry optimizations were carried out using the Broyden-Fletcher-Goldfarb-Shanno algorithm [35-38]. Both atomic positions and in-plane lattice parameters were relaxed using a variable cell relaxation scheme with two-dimensional in-plane cell freedom, while the out-of-plane lattice parameter was kept fixed to preserve the vacuum separation. The relaxation process was continued until the residual forces on all atoms were below $9.72 \times 10^{-4}$ Ry/Bohr. After relaxation, the optimized structures were analyzed in terms of C-C and C-N bond lengths around the defect region. To quantify the degree of local structural distortion and symmetry breaking, the bond-length variations were defined as:

$$d_{C-C} = d_{C-C}{}^{\max} - d_{C-C}{}^{\min} \quad (1)$$

$$d_{C-N} = d_{C-N}{}^{\max} - d_{C-N}{}^{\min} \quad (2)$$

where $d^{max}$ and $d^{min}$ represent the maximum and minimum bond lengths within the selected bonding environment. These descriptors were used to compare the local reconstruction behavior of the peripheral N configurations: pyridinic, pyridazinic, pyrrolic, and pyrazolic. The optimized structures were post-processed and visualized using Python. To further identify charge accumulation and depletion induced by nitrogen incorporation, the charge-density difference was calculated as:

$$\Delta\rho = \rho_{defective} - \rho_{defective\,framework} - \sum \rho_N \quad (3)$$

where $\rho_{defective}$ is the charge density of the complete nitrogen-containing defective graphene system, $\rho_{defective\,framework}$ is the charge density of the corresponding defective carbon framework, and $\sum \rho_N$ represents the charge density contribution from the introduced nitrogen atoms. All charge densities were evaluated using the same supercell and atomic positions to ensure meaningful subtraction.

Bader charge analysis was performed to quantify the atomic charge redistribution. The total charge densities were exported as volumetric cube files using the Quantum ESPRESSO post-processing tool pp.x [29, 30] and then partitioned into atomic basins using the Henkelman group's Bader analysis algorithm [39]. The atomic charge transfer was defined as:

$$\Delta q_i = q_i^{Bader} - q_i^{Neutral} \quad (4)$$

where $q_i^{Bader}$ is the Bader charge associated with atom i, and $q_i^{Neutral}$ is the reference valence charge of the corresponding neutral atom.

The computational protocol was validated by calculating a standard substitutional nitrogen-doped graphene model before analyzing the peripheral nitrogen configurations. Convergence tests were performed for the plane-wave cutoff energy, charge-density cutoff, and k-point sampling before production calculations. The final computational parameters were selected such that further increases in cutoff energy or k-point density produced negligible changes in the total energy. The same converged parameters were then used for all structural, electronic, charge-density, and magnetic-property calculations.

***Defect formation energy calculation***

The relative thermodynamic stability of the peripheral nitrogen configurations was evaluated using the defect formation energy, $\Delta E_f$, following the standard chemical-potential-based formalism for first-principles defect calculations [40, 41], calculated as:

$$\Delta E_f = E_{defective} - E_{pristine} + n_C \times \mu_C - n_N \times \mu_N \tag{5}$$

where $E_{defective}$ is the total energy of the relaxed defective graphene system, and $E_{pristine}$ is the total energy of the pristine 5×5 graphene supercell. The terms $n_C$ and $n_N$ denotes the number of carbon atoms removed and nitrogen atoms added, respectively. The carbon chemical potential, $\mu_C$ was taken as the total energy per carbon atom in pristine graphene, while the nitrogen chemical potential, $\mu_N$ was obtained from half the total energy of an isolated $N_2$ molecule calculated using the same exchange-correlation functional and pseudopotential settings.

***Electronic structure analysis***

The electronic band structures were calculated after self-consistent convergence using the optimized geometries. Band dispersions were evaluated along the standard graphene high-symmetry path: Γ → K → M → Γ [2].

Spin-up and spin-down channels were calculated separately for all spin-polarized systems. The Fermi level was set to zero energy, and the band energies were plotted as $E\text{-}E_f$. Any band gap obtained from the GGA-PBE calculations was interpreted as a qualitative PBE-level electronic trend rather than an exact experimental band gap, since semilocal functionals commonly underestimate band gaps [31].

***Magnetic analysis***

The total density of states and spin-polarized projected density of states were calculated using a denser 14×14×1 [34] Monkhorst-Pack k-point mesh in non-self-consistent calculations. Orbital-projected contributions were obtained using the projwfc.x module [29, 30] . For spin-polarized DOS plots, spin-up states were plotted on the positive axis and spin-down states on the negative axis, with the Fermi level aligned at 0 eV. To examine defect-induced magnetism, all defective graphene configurations were calculated using spin-polarized DFT [42]. The total magnetization, M, was obtained from the converged spin-polarized charge density according to:

$$M = \int_{\Omega} [\rho_{\uparrow}(\mathbf{r}) - \rho_{\downarrow}(\mathbf{r})] \, d\mathbf{r} \tag{6}$$

where $\rho_{\uparrow}(\mathbf{r})$ and $\rho_{\downarrow}(\mathbf{r})$ are the spin-up and spin-down electron densities, respectively.

The spatial spin-density distribution was calculated as:

$$\rho_s(r) = \rho_{\uparrow}(\boldsymbol{r}) - \rho_{\downarrow}(\boldsymbol{r}) \tag{7}$$

Spin-density maps were generated using the Quantum ESPRESSO post-processing tool [29] and visualized using VESTA [43].

## Results and Discussion

### *Validation of the Computational Framework*

Before analyzing the peripheral nitrogen configurations, the reliability of the computational framework was evaluated using a standard substitutional nitrogen-doped graphene model. The optimized substitutional N-doped graphene structure is shown in Fig. 1(a). Full structural relaxation results in the overall honeycomb structure of graphene being maintained and the nitrogen dopant being substitutionally incorporated into the carbon lattice. The absence of significant lattice disruption allows for the conclusion that the adopted supercell, relaxation procedure, and convergence criteria are appropriate for the description of nitrogen incorporation in graphene-based systems. This optimized structure was then used to benchmark the energetic and magnetic response of the computational model. The calculated defect formation energy for the substitutional N-doped graphene is 0.068028 Ry. It is close to the previously reported values of 0.066 Ry [21] and ~0.066885 Ry [44] for similar-sized substitutional N configurations.

The slight deviation from the reported values is within the expected range due to the possible differences in pseudopotential type, exchange-correlation treatment, k-point sampling and relaxation criteria. The agreement verifies that the present computational settings can reliably reproduce the relative energetic cost of nitrogen incorporation into graphene.

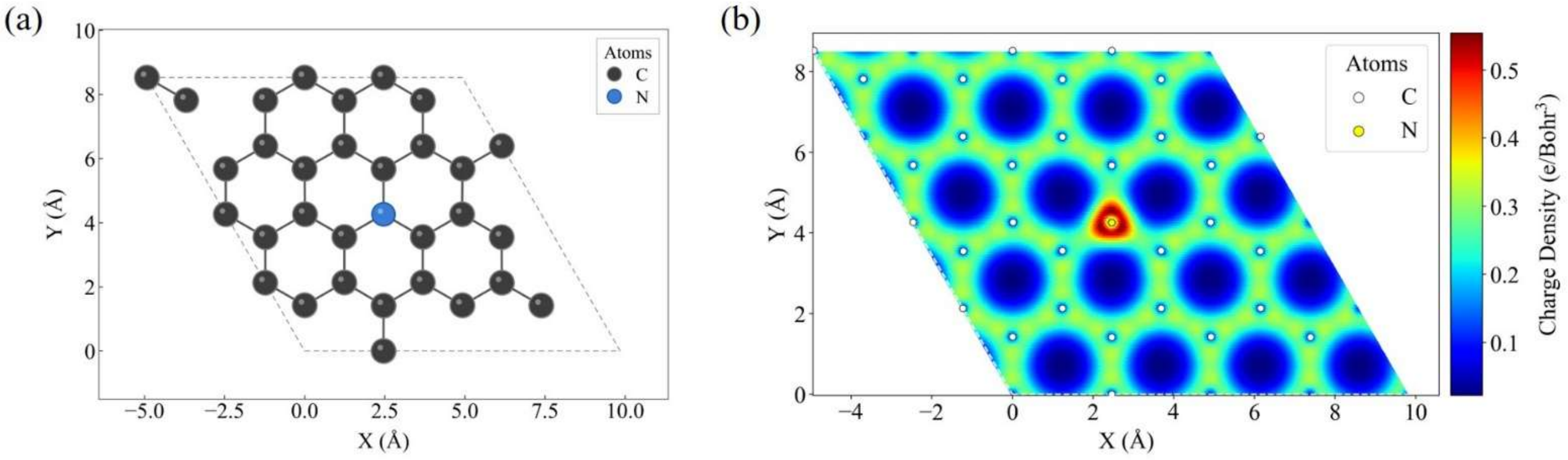

**Figure 1.** (a) Optimized structure, and (b) Charge-density map of substitutional N-doped graphene

In addition to the formation energy, the magnetic response of the substitutional N-doped graphene system was examined. The calculated total magnetization is 0.00 $\mu_B$, indicating a nonmagnetic ground-state configuration for isolated substitutional nitrogen in graphene under the present computational conditions. This result is consistent with previous reports showing that substitutional or graphitic nitrogen defects in graphene exhibit zero net magnetization [45, 46]. The absence of net magnetization is physically reasonable because substitutional nitrogen mainly contributes an extra electron to the delocalized π-electron network without creating a dangling-bond state.

**Table 2:** Numerical validation against previous studies.

| **Property** | **Current Study** | **Previous Study** |
| --- | --- | --- |
| Defect formation energy (Ry) | 0.068028 | 0.066 [21]; 0.066885 [44] |
| Total magnetization ($\mu_B$) | 0.00 | 0.00 [45]; 0.00 [46] |

The charge density distribution of the model, shown in Fig. 1(b), further supports the expected electronic behavior of substitutional nitrogen in graphene. The charge density is locally modified around the nitrogen dopant site, reflecting the higher electronegativity and different valence-electron configuration of nitrogen compared with carbon. However, this local charge redistribution does not produce a finite total magnetic moment. This validation provides a solid basis for extending the same computational framework to the more complex peripheral nitrogen configurations investigated in the present work. Unlike simple substitutional nitrogen, the pyridinic, pyridazinic, pyrrolic and pyrazolic systems discussed here have nitrogen atoms around a central void in the graphene system. Thus, their structure, charge redistribution, electronic states, and magnetic properties can be anticipated to be determined not only by nitrogen incorporation but also by the topology and bonding geometry of the vacancy edge.

***Structural Properties***

The optimized geometries of four peripheral N configurations are shown in Fig. 2. The present systems differ from conventional substitutional nitrogen doping in which an individual carbon atom is replaced by nitrogen in the basal plane of graphene, but instead involve nitrogen atoms at the reconstructed edge of a central graphene void. The reconstructed edges are not identical in the four systems, due to the different peripheral positions of the nitrogen atoms and the different local C-N coordination environment. All these geometric differences should affect the local strain distribution, the continuity of $\pi$-electrons, the localization of defect-states, and the magnetic response of the defective graphene lattice. The C-C bond lengths and the C-N bond lengths were analyzed after fully relaxing the structure in order to quantify the structural reconstruction. The minimum, maximum, average, and variation of the corresponding values are presented in Table 3. The bond-length variation serves as a good indicator of the distortion and symmetry breaking around the void.

There is a significant configuration dependence of the C-C bond lengths. It is found that pyridazinic N-doped graphene has the highest C-C bond-length distortion among the four systems, where the C-C bond-lengths vary from 1.358 to 1.761 Å, and the variation is 0.403 Å. This significant distortion suggests significant reconstruction of the carbon framework locally near the void boundary. This bond lengthening of the C-C bond in the pyridazinic configuration indicates that the local carbon network is under significant strain, likely because of the asymmetric position of the two adjacent N atoms and the defect-edge bonding network reorganization. The weak covalent bond formation between the unpaired electrons of the C1 and C2 atoms was also responsible for the symmetry breaking. As seen in Fig. 2(b), the two atoms came closer, at 1.649 Å. Such interactions led to the formation of a pentagonal ring, initiating the Jahn-Teller distortion.

The same was seen in other configurations, as seen in (a, c-d) in Fig. 2. But in the case of the C3-C4 bond adjacent to the void, the bond length extends to 1.761 Å, which is significantly larger than that of pristine graphene. The elongation is due to both the void-edge reconstruction and the large electronic perturbation caused by the neighboring pyridazinic N atoms. The optimized structure is well converged and has a finite formation energy, even though the bond length is larger than expected, the defect is not dissociative, but rather a metastable configuration. The highly elongated C–C bond can be interpreted as a localized covalent strain-relief bridge rather than as a fully dissociated bond. In this reconstruction pathway, the defective lattice avoids a more extensive global distortion or out-of-plane buckling by localizing the tensile strain into a specific edge bridge near the vacancy. This localized bond sacrifice stabilizes the reconstructed pore edge and provides a direct structural signature of topology-driven strain accommodation in the nitrogen-decorated graphene lattice.

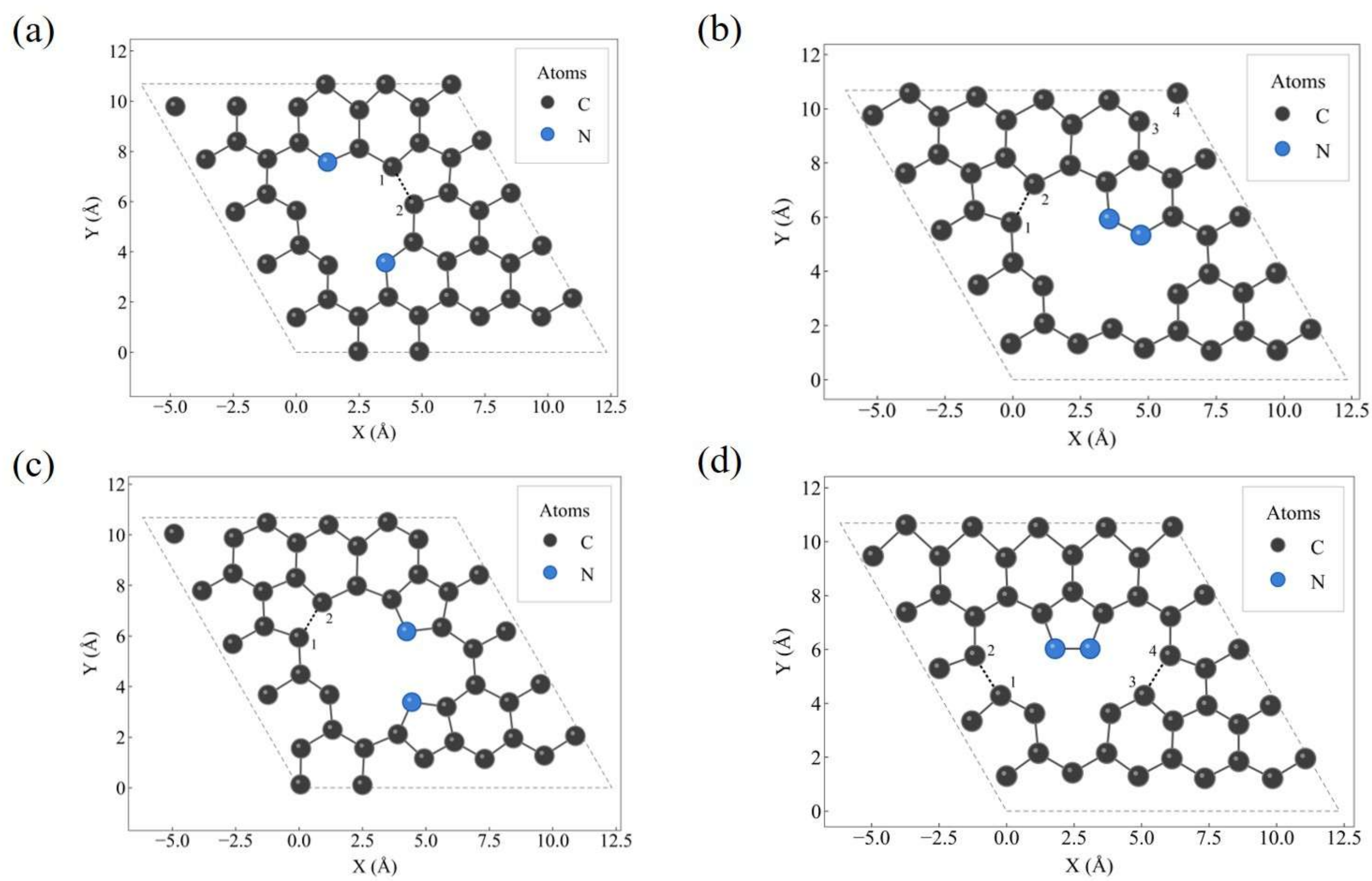

**Figure 2.** Optimized structures of the four peripheral N configurations: (a) pyridinic, (b) pyridazinic, (c) pyrrolic, and (d) pyrazolic.

There is also a major C-C distortion in the pyrazolic configuration, with C-C bond lengths varying between 1.362 and 1.698 Å and a difference of 0.336 Å. This distortion is less than in the pyridazinic system, but larger than in the pyridinic and pyrrolic systems. This indicates that, despite the high symmetry of the two C-N bonds in this topology, the pyrazolic topology itself is strained with respect to the surrounding carbon network. The coexistence of the strong C-C distortion and uniform C-N bonding is an important structural feature of the pyrazolic system and may be related to the unique electronic and magnetic behavior to be discussed later. An intermediate level of C-C distortion is found in the pyridinic configuration. Its C-C bond lengths vary from 1.357 to 1.675 Å, giving a variation of 0.318 Å. This implies that the pyridinic peripheral structure provides a relatively good balance between defect-edge reconstruction and N-stabilization. By contrast, the pyrrolic configuration has the least variation in C-C bond length, with values of 1.361 to 1.572 Å and a difference of only 0.211 Å. Lower C-C distortion implies that the carbon framework can relax more smoothly around the void and, thus, the elongation of the carbon bond and the structural strain in pyrrolic topology is also less.

**Table 3:** C-C and C-N bond-length statistics for pyridinic, pyridazinic, pyrrolic, and pyrazolic configurations.

| System | Bond length (Å) | | | | | | | | $\Delta E_f$ (Ry) |
|---|---|---|---|---|---|---|---|---|---|
| | C-C | | | | C-N | | | | |
| | ${d_{C-C}}^{min}$ | ${d_{C-C}}^{max}$ | avg | $d_{C-C}$ | ${d_{C-C}}^{min}$ | ${d_{C-C}}^{max}$ | avg | $d_{C-C}$ | |
| Pyridinic | 1.357 | 1.675 | 1.516 | 0.318 | 1.368 | 1.379 | 1.373 | 0.011 | 0.82954397 |
| Pyridazinic | 1.358 | 1.761 | 1.444 | 0.403 | 1.362 | 1.374 | 1.368 | 0.012 | 1.08275759 |
| Pyrrolic | 1.361 | 1.572 | 1.467 | 0.211 | 1.382 | 1.405 | 1.393 | 0.023 | 0.95990110 |
| Pyrazolic | 1.362 | 1.698 | 1.530 | 0.336 | 1.398 | 1.398 | 1.398 | 0.000 | 1.19121619 |

Additional information on the local N bonding environments is gained from the C-N bond length distribution. The pyridazinic configuration has the lowest average C-N bond length of 1.368 Å with the pyridinic configuration having an average C-N bond length of 1.373 Å. The relative shortness of these C-N bonds suggests greater local C-N interaction, as well as more tightly bound nitrogen in the defect boundary. The small C-N bond-length variations in pyridinic and pyridazinic systems, 0.011 and 0.012 Å, respectively, further suggest that the nitrogen atoms form relatively well-defined bonding environments despite the surrounding carbon-framework reconstruction. The pyrrolic configuration exhibits a longer average C-N bond length of 1.393 Å and the largest C-N bond-length variation, 0.023 Å. This indicates that the two nitrogen sites in the pyrrolic configuration are not equivalent in their local bonding environment. Such C-N asymmetry may affect local charge redistribution and spin localization, because non-equivalent nitrogen sites can disturb the surrounding π-electron network differently. The pyrazolic configuration is structurally distinctive because both C-N bonds are identical, with a C-N bond length of 1.398 Å. This perfect C-N uniformity suggests a highly symmetric nitrogen coordination environment. However, the pyrazolic system also has the longest average C-N bond length among the four configurations. The longer but perfectly uniform C-N bonds indicate that the pyrazolic motif forms a symmetric, but less tightly bound, nitrogen-edge configuration. This structural symmetry is particularly important because the pyrazolic system is later found to exhibit electronic and magnetic responses that differ from those of the other three systems.

The relative energetic stability of the four nitrogen configurations was evaluated using defect-formation energies. The calculated formation energies follow the order: Pyridinic < Pyrrolic < Pyridazinic < Pyrazolic, with values of 0.82954397, 0.95990110, 1.08275759, and 1.19121619 Ry for pyridinic, pyrrolic, pyridazinic, and pyrazolic configurations, respectively. This energetic sequence identifies pyridinic N-doped graphene as the most thermodynamically favorable peripheral nitrogen configuration among the four systems considered. The lower formation energy of the pyridinic configuration indicates that the graphene vacancy edge can accommodate this nitrogen arrangement with relatively lower energetic cost. The pyrrolic configuration has the second-lowest formation energy, despite showing the smallest C-C bond distortion. This indicates that the energetic stability of peripheral N-doped graphene depends not only on low carbon-framework distortion, but also on other factors. Rather, the stability is achieved by a complex interplay among the reconstruction of the carbon edges, the strength of the C-N bond, and the coordination of N. The pyridazinic configuration has a higher formation energy than both pyridinic and pyrrolic systems. This is in accordance with a large C-C bond-length distortion, particularly the presence of the longest C-C bond among the four systems. The strong local strain associated with the pyridazinic topology may account for the higher energetic cost. The pyridazinic structure does, however, have relatively short C-N bonds, suggesting that high formation energy is largely due to strain in the reconstructed carbon structure and not only to weak C-N bonding. The pyrazolic configuration has the maximum formation energy, is the least thermodynamically favorable of the four peripheral nitrogen motifs explored. This is interesting because the C-N bonds are perfectly equal in this pyrazolic structure, and yet it is energetically costly. Therefore, structural symmetry in the C-N bonds does not necessarily guarantee energetic stability. The longer C-N bonds, significant C-C distortion, and the particular local topology are more likely to be associated with the higher formation energy of the pyrazolic system.

***Electronic Properties***

Since the four systems contain the same parent graphene framework, while the peripheral arrangements of the N atoms differ, the direct comparison reveals how nitrogen topology controls electron localization, defect-state formation, and band-gap modulation. The total charge-density maps of the four optimized configurations are shown in Fig. 3. For all systems, a charge depleted region is seen in the void center, and increased charge localization is seen around the edge of the void containing N. This redistribution is anticipated as nitrogen is more electronegative than carbon, and is the reason for the shift in electron density in the carbon framework. The relative position and bonding topology around the vacancy, however, is also crucial to the electronic response as demonstrated by the spatial pattern of charge localization in the four nitrogen motifs. For the pyridinic configuration, the charges are localized at the two spatially distributed N sites at defect boundary. The areas around the N atoms are highly charged, with the central void being highly depleted. This suggests that the pyridinic edge configuration distorts the electronic density locally, but does not completely disrupt the extended graphene framework. The spatial separation of the two nitrogen centers also suggests that each N site perturbs a different part of the vacancy rim, giving rise to a distributed charge-reorganization pattern around the vacancy. The two N sites have almost identical negative charge transfer values of -2.55 and -2.58.

This indicates that both pyridinic N atoms behave as effective accumulation centers and participate similarly in polarizing the surrounding carbon framework. The nearly balanced N-site charge transfer is consistent with the distributed charge-density pattern observed in the total charge-density and charge-density-difference maps. The pyridazinic configuration displays a more localized and adjacent charge accumulation pattern because the two nitrogen atoms are placed closer to each other along the vacancy edge. The high-density region is therefore concentrated on one side of the void rather than distributed over opposite edge positions. This indicates that neighboring N atoms act cooperatively as an electron-rich defect center. The localized nature of the charge density around the pyridazinic N pair is consistent with its strong local structural distortion discussed in the previous section. The pyridazinic configuration also shows comparable values of charge transfer -1.29 and -1.27 on the adjacent nitrogen atoms, but the magnitude is lower than that of the pyridinic case. This suggests that although the adjacent N pair forms a localized electron-rich region, the charge accumulation per N atom is moderated by the close N-N arrangement and shared interaction with the neighboring carbon network.

In the pyrrolic configuration, again, the charge density is increased around the defect edge, but the distribution is not as symmetric as in the pyridinic configuration. The two N sites are not equally contributing to the local electronic density surrounding the atoms, suggesting local electronic non-equivalence as a result of the pyrrolic geometry. This feature is significant because it results in non-uniform charge transfer, local electronic defect states, and spin-polarized electronic features near the Fermi level. The pyrrolic configuration is electronically unique, with the two N sites having very different charge accumulation of -1.07 and -2.61. The large charge asymmetry in the N site is a sign of the non-equivalence of the two N atoms in the pyrrolic defect topology. This is an extra degree of asymmetry in charge transfer that gives an important electronic support to the localized defect states and the spin-polarized features of the band structure and density of states.

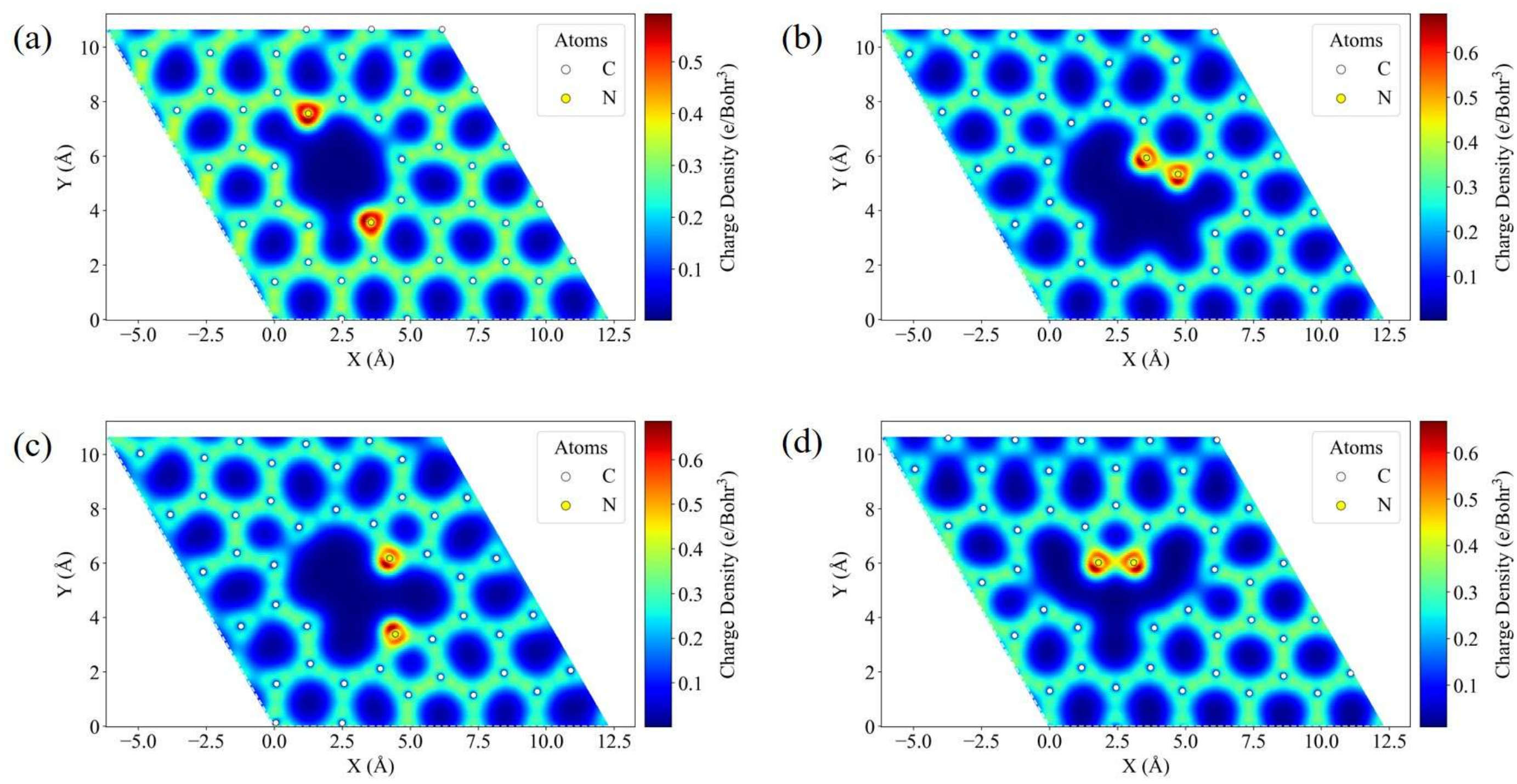

**Figure 3.** Total charge-density maps of (a) pyridinic, (b) pyridazinic, (c) pyrrolic, and (d) pyrazolic configurations.

The pyrazolic structure exhibits a relatively evenly distributed charge-localization pattern around the adjacent nitrogen pair. This is in agreement with the structural result, which indicates equal lengths of C-N bonds in the pyrazolic system. Hence, the charge-density map indicates that the pyrazolic motif creates a more even electronic environment around the N pair, which is subsequently manifested in the unique band-gap opening and spin-compensated character of the pair. It displays almost equal charge transfer between the two nitrogen atoms of -1.22 and -1.26. Thus, despite having the highest formation energy among the four configurations, the pyrazolic system is the most electronically compensated configuration.

The charge-density difference maps were analyzed to differentiate simple localization of an electron from actual charge-redistribution due to the introduction of nitrogen. The charge-density difference plots are shown in Fig. 4. These maps show the areas of electron accumulation and depletion that result from the interaction of the nitrogen atoms and the defective graphene framework. The charge redistribution around the vacancy rim is clearly seen in all four systems, which shows that a significant change in the local electronic structure of defective graphene can be introduced by the incorporation of peripheral nitrogen atoms. The charge accumulation and depletion lobes are mainly concentrated near the C-N bonds and neighboring carbon atoms, indicating strong bond polarization at the defect boundary. This confirms that the peripheral N atoms actively reshape the local electronic landscape of the vacancy edge.

In the pyridinic configuration, charge redistribution occurs around both nitrogen sites and extends into the neighboring carbon framework. The spatially separated accumulation and depletion regions indicate that the pyridinic N atoms have a relatively large effect on the π-electron network of the graphene sheet. This extended redistribution can help to form defect-derived electronic states near the Fermi level. In the pyridazinic configuration, there is a higher concentration of charge redistribution around the adjacent N atoms. The two neighboring N atoms are close together, so the electronic perturbations of the two atoms overlap, leaving a compact and intense redistribution region at one side of the vacancy. This local charge rearrangement is in line with the high C-C bond distortion noticed for this configuration, and suggests that there exists a strong correlation between the structural reconstruction and the electronic polarization.

The pyrrolic configuration shows an asymmetric pattern of the charge-density difference, indicating unequal electronic involvement of the two N sites. This asymmetry is crucial as it can result in non-equivalent local potential environments around the vacancy. This can influence the position and occupancy of defect states near the Fermi level. The pyrazolic configuration displays a relatively even distribution around the adjacent nitrogen pair. In the case of charge accumulation and depletion, the pattern is more symmetric with respect to the N-N centered region than in the pyridazinic case. The identical C-N bond lengths are indicative of a higher degree of electronic compensation and are consistent with such a balanced redistribution.

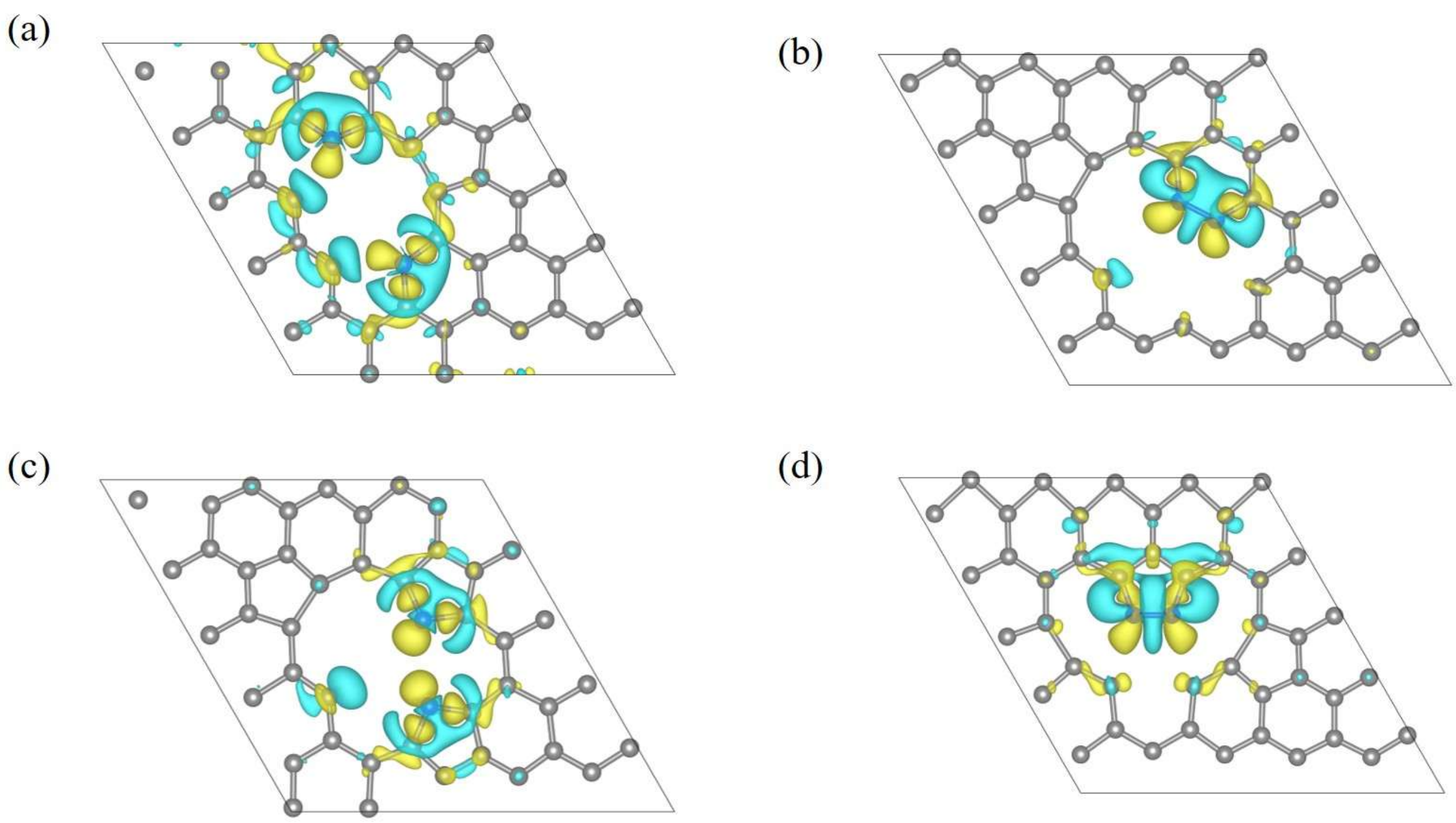

**Figure 4.** Charge-density difference plots of (a) pyridinic, (b) pyridazinic, (c) pyrrolic, and (d) pyrazolic configurations. The yellow and cyan isosurfaces represent electron accumulation and depletion, respectively, plotted at an isosurface level of 0.004 e/Å$^3$

The spin-resolved electronic band structures of the four configurations are shown in Fig. 5. The Fermi level is set to 0 eV in all cases. The comparison shows the topology-dependent modulation of the electronic structure of N-doped defective graphene. The pyridinic configuration shows electronic states that are near the Fermi level, suggesting a metallic or near-metallic state. Bands around the $E_f$ are observed, indicating that pyridinic edge N creates defect-derived states in the low-energy electronic window of graphene. These states are the result of the reconstruction of the vacancy boundary and the hybridization of N 2p orbitals with the neighboring C 2p network. The absence of a clear band gap indicates that the pyridinic motif preserves electronic conductivity while introducing localized defect states. The pyridazinic configuration also shows bands that cross the Fermi level, confirming its metallic or near-metallic character. The band dispersion suggests that the pyridazinic defect modifies the graphene electronic structure without opening a significant global gap. The pyrrolic configuration similarly retains low-energy electronic states around the Fermi level. Although the pyrrolic system shows the smallest C-C bond distortion among the four configurations, its asymmetric C-N bonding and uneven charge transfer produce defect-related electronic states near $E_f$. In marked contrast, the pyrazolic configuration exhibits a clear band gap of 0.377 eV at the PBE level. While pyridinic, pyridazinic, and pyrrolic configurations behave as metallic or near-metallic systems, the pyrazolic arrangement converts the defective graphene system into a narrow-gap semiconducting state. The pyrazolic result demonstrates that peripheral N topology can be used as a chemical design parameter to induce band-gap opening in defective graphene.

The origin of this band-gap opening can be related to the structural and charge-transfer features discussed above. In this topology, the two nitrogen atoms form a more electronically balanced pair, producing identical C-N bond lengths and nearly symmetric charge redistribution around the vacancy rim. This symmetry promotes effective passivation of dangling vacancy-edge states and suppresses uncompensated mid-gap defect bands that remain active in the other systems. As a result, the occupied and unoccupied defect-derived states are separated around the Fermi level, producing the observed narrow PBE-level semiconducting gap. Thus, the pyrazolic configuration behaves as an electronic switch: it transforms a vacancy-containing N-doped graphene system from a metallic/near-metallic state into a narrow-gap semiconducting state.

To identify the orbital origin of the electronic states near the Fermi level, spin-polarized projected density of states were analyzed. The total and orbital-projected DOS plots are shown in Fig. 7. The projected states include contributions from C 2s, C 2p, and N 2p orbitals. Across all configurations, the C 2p states dominate the electronic structure over a broad energy range, confirming that the graphene π-network remains the principal electronic framework. However, the N 2p orbitals contribute significantly, especially near the Fermi level. This indicates that the electronic properties of the systems arise from hybridization between the N 2p states and neighboring C 2p orbitals around the defect boundary.

For the pyridinic configuration, the DOS near the Fermi level contains contributions from both C 2p and N 2p states. This confirms that the low-energy defect states are not purely carbon-derived. Rather, they originate from C-N hybridization at the reconstructed vacancy edge. The presence of finite DOS near $E_f$ supports the metallic or near-metallic behavior observed in the band structure. The pyridazinic configuration shows sharper orbital features associated with the localized N pair and surrounding carbon atoms. These features are consistent with the compact charge redistribution and strong structural distortion of this configuration. The N 2p contribution near the low-energy region indicates that the adjacent nitrogen atoms actively participate in forming the defect-state spectrum. In the pyrrolic configuration, the DOS reflects the electronic asymmetry of the two nitrogen sites. The N 2p contribution appears strongly localized in selected energy regions, indicating that the pyrrolic N atoms do not contribute equivalently to the electronic structure. This agrees with the large N-site charge asymmetry observed in the Bader analysis and supports the conclusion that pyrrolic graphene has a locally non-equivalent electronic environment. For the pyrazolic configuration, the DOS near the Fermi level is reduced compared with the metallic/near-metallic configurations, consistent with the band gap observed in the band structure. The orbital-resolved DOS suggests that the paired and electronically compensated N atoms modify the defect-state spectrum in such a way that occupied and unoccupied states become separated around $E_f$. This confirms that pyrazolic peripheral nitrogen topology is uniquely capable of opening a band gap in the defective graphene lattice.

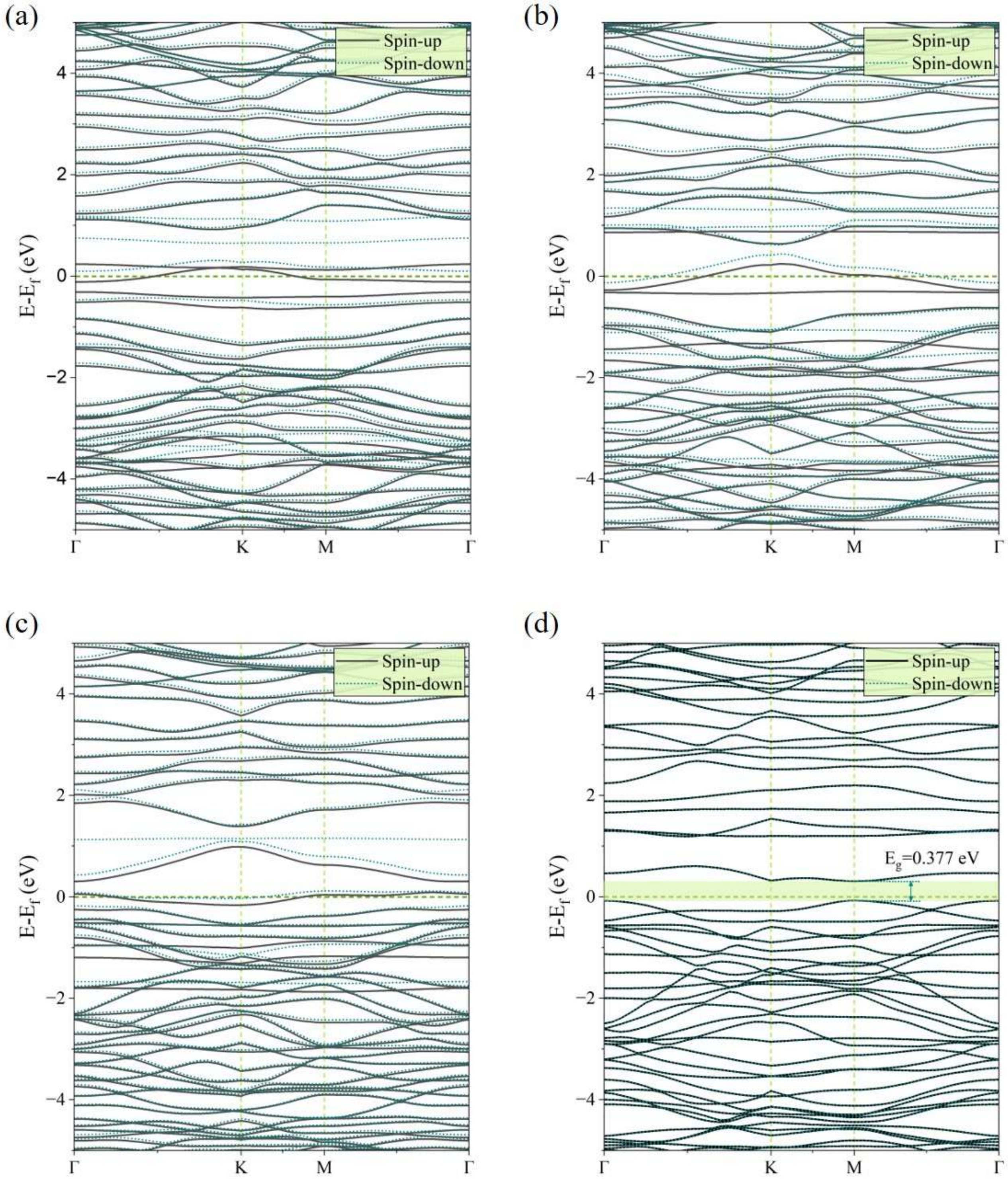

**Figure 5.** Spin-resolved band structures of (a) pyridinic, (b) pyridazinic, (c) pyrrolic, and (d) pyrazolic configurations

The N-atom-resolved SPDOS provides deeper insight into the orbital character of the nitrogen-induced electronic states. The decomposed N 2p contributions are shown in Fig. 8, where the $2p_x/2p_y$ components represent in-plane bonding contributions and the $2p_z$ component is associated with the out-of-plane π-electronic network of graphene. The decomposition shows that nitrogen-induced electronic states arise from both in-plane and out-of-plane orbital channels. The in-plane $2p_x/2p_y$ orbitals contribute to local C-N bonding at the defect rim, while the $2p_z$ states interact with the graphene π-network and therefore have a stronger role in controlling states near the Fermi level. In the metallic or near-metallic pyridinic, pyridazinic, and pyrrolic configurations, the N 2p-derived states participate in the low-energy electronic structure near the Fermi level. This confirms that the peripheral N atoms are directly involved in creating defect states rather than merely causing a passive electrostatic perturbation. The reduction of the density of states at the Fermi level in the pyrazolic configuration, with a more balanced electronic environment, contributes to the opening of the PBE-level band gap.

***Magnetic Properties***

All the pyridinic, pyridazinic and pyrrolic configurations yield finite magnetization while the pyrazolic configuration gives zero net magnetization. The total magnetization of pyridazinic N-doped graphene is the highest at 1.49 μB/cell, followed by pyridinic, and pyrrolic at 1.35 μB/cell and 1.22 μB/cell, respectively. The pyrazolic configuration, on the other hand, has full net spin compensation. This is in line with the electronic property analysis, in which the pyrazolic configuration was found to open a PBE-level band gap of 0.377 eV.

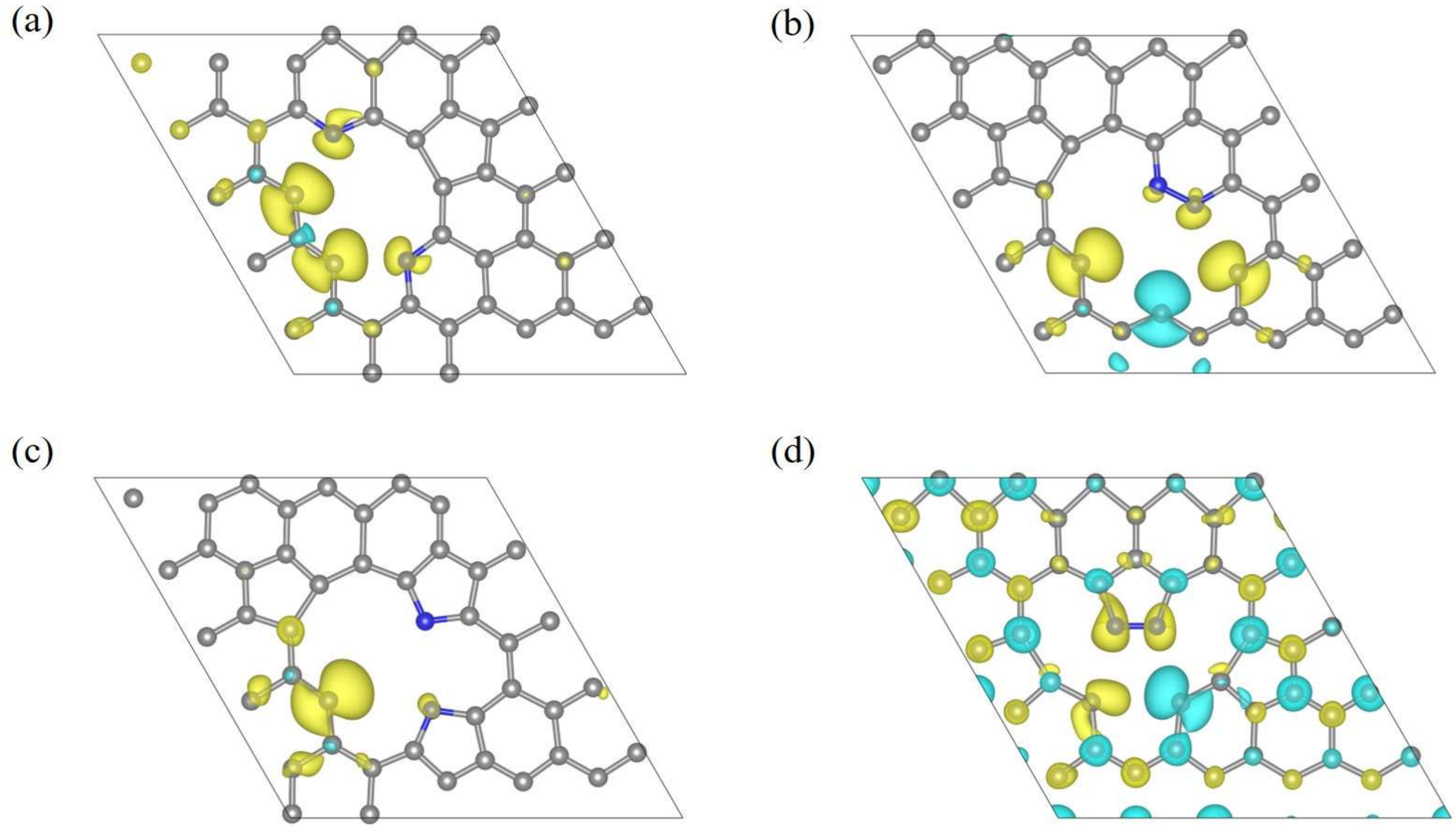

**Figure 6.** Spin-density plots of (a) pyridinic, (b) pyridazinic, (c) pyrrolic, and (d) pyrazolic configurations. The yellow and cyan isosurfaces represent spin-up and spin-down charge densities, respectively, plotted at an isosurface level of 0.005 e/Å³ (a-c), and $2\times10^{-5}$ (d).

The spin-density plots are shown in Fig. 6. In the pyridinic configuration, the spin density is concentrated at the boundary of the defect and over neighboring carbon atoms. It means that the finite magnetic moment is not solely due to the N dopants. Rather, the peripheral N atoms alter the local electronic structure of the vacancy edge, giving rise to the presence of spin-polarized states that spread out over both N and neighboring C sites. This behavior is characteristic of defect-induced magnetism in graphene, where vacancy-edge carbon atoms and the dopant-modified π-states can both contribute to the net spin polarization. For the pyridazinic structure, a similar localized behavior is found with a more pronounced asymmetry in the spin density and a larger contribution from the neighboring atoms, which indicates a higher degree of orbital hybridization around the defect site. The strong positive and negative spin-density lobes suggest that there is high spin polarization and significant unpaired electron redistribution. In the pyrrolic configuration, the spin density is more localized on fewer carbon atoms and thus localized magnetic states are created. This results in a smaller net magnetic moment in spite of the presence of localized unpaired spins. The pyrazolic form, on the other hand, has the widest-spread spin-density distribution, which extends over a larger part of the lattice and involves both N and C atoms. The coexistence of comparable spin-up and spin-down contributions results in complete spin compensation, yielding a nonmagnetic ground state.

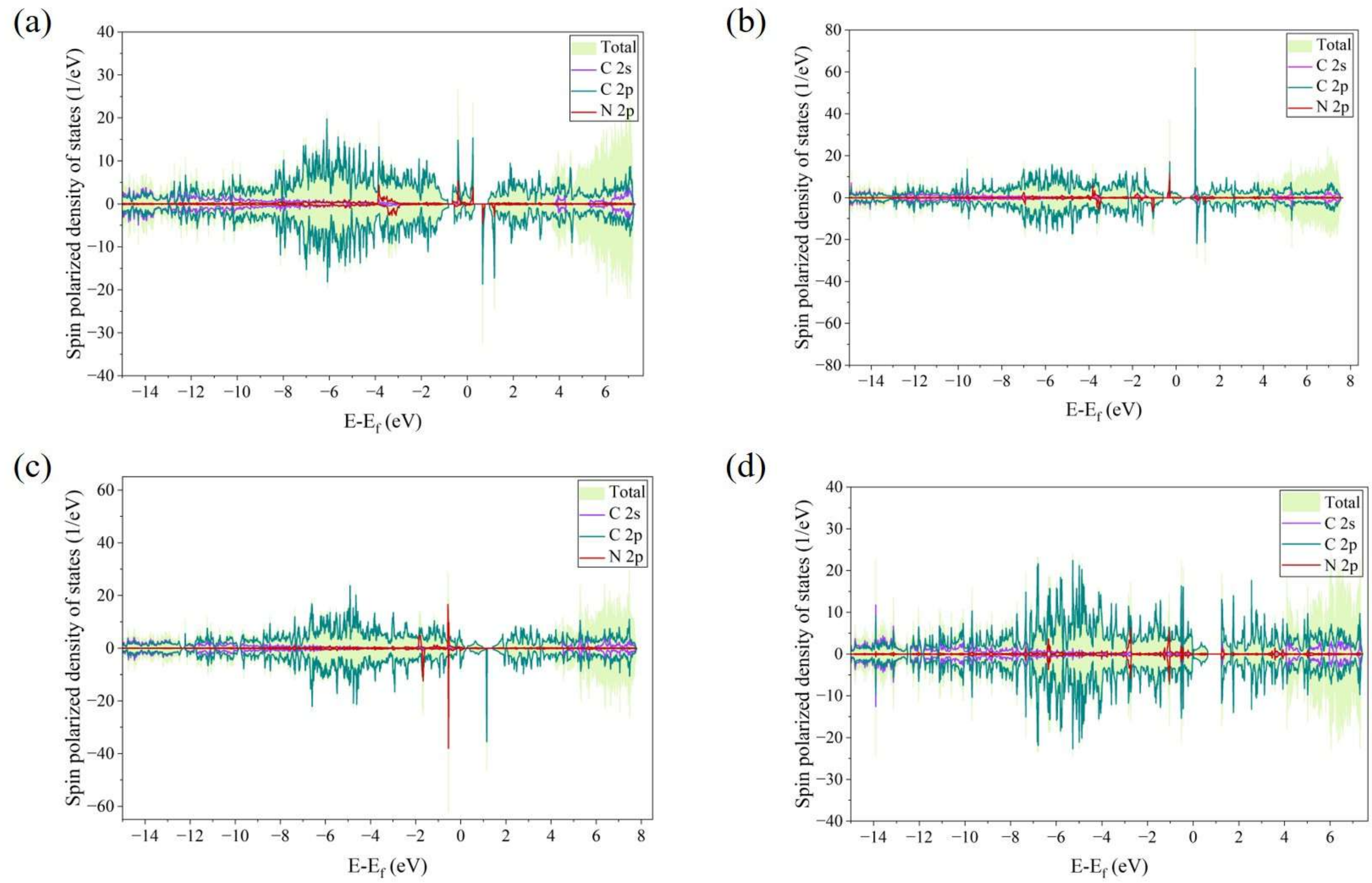

**Figure 7.** Total and orbital-projected SPDOS of (a) pyridinic, (b) pyridazinic, (c) pyrrolic, and (d) pyrazolic configurations.

In the pyridinic, pyridazinic, pyrrolic configurations the maximum local magnetic moments are localized on selected C atoms at the rim of the vacancy. Both nitrogen atoms and neighboring carbon atoms contribute, but the surrounding carbon atoms often play a major role in carrying the local spin density. The spin-polarized projected density of states provides further insight into the electronic origin of the observed magnetic behavior. The total and orbital-projected SPDOS plots are shown in Fig. 7. For the pyridinic, pyridazinic, and pyrrolic configurations, the spin-up and spin-down density of states are not identical, confirming that these systems possess spin-polarized electronic structures. The presence of spin-asymmetric states near the Fermi level is particularly important because these states determine the net magnetic moment. The orbital-projected DOS indicates that the relevant states are dominated by C 2p contributions, with additional participation from N 2p states. This confirms that the magnetic response is mainly associated with the defect-modified π-electron network of graphene.

The pyridinic configuration shows spin-asymmetric electronic states around the low-energy region, consistent with its finite magnetic moment of 1.35 $\mu_B$. The pyridazinic configuration shows the strongest total magnetization, and its SPDOS reflects pronounced spin imbalance in the defect-related electronic states. The adjacent nitrogen pair produces localized orbital features that interact with neighboring carbon 2p states. This local electronic structure supports stronger spin polarization and explains why pyridazinic graphene has the highest magnetic moment among the four configurations. The pyrrolic configuration also shows spin-up/spin-down asymmetry, but with a slightly smaller total magnetization than pyridinic and pyridazinic systems. For the pyrazolic configuration, the spin-polarized DOS is consistent with zero net magnetization and band-gap formation. The reduction of states at the Fermi level and the more balanced spin response indicate that the pyrazolic motif suppresses uncompensated spin-polarized defect states.

Therefore, peripheral nitrogen doping around graphene vacancies provides a chemical means to select between magnetic and nonmagnetic electronic states. Pyridinic, pyridazinic, and pyrrolic motifs produce finite spin-polarized magnetic moments, whereas the pyrazolic motif generates a spin-compensated nonmagnetic state with a narrow PBE-level band gap. Therefore, by controlling the local heterocyclic-like nitrogen arrangement around a graphene void, one can tune graphene from a magnetic metallic/near-metallic defect system to a nonmagnetic semiconducting system.

To clarify the role of nitrogen orbitals in magnetic-state formation, the N 2p orbital-resolved SPDOS was analyzed. Since magnetism in graphene-based defects is generally associated with spin polarization of $p_z$-derived π states, the N $2p_z$ contribution is especially important for identifying how peripheral N atoms tune the magnetic response. For the pyridinic, pyridazinic, and pyrrolic configurations, the N 2p-derived states contribute to the spin-polarized electronic structure near the Fermi level. These contributions indicate that nitrogen atoms are directly involved in modifying the spin-dependent defect states. However, the total magnetic behavior cannot be assigned to N atoms alone, because the total SPDOS and spin-density maps show strong participation of neighboring C 2p states. Thus, nitrogen acts as a spin-modulating center that polarizes the defect-edge carbon network. The pyrazolic configuration shows a different N-orbital response. The more compensated N pair reduces spin imbalance near the Fermi level. This suggests that the pyrazolic N arrangement reorganizes the N 2p and C 2p hybridized states in a way that suppresses net spin polarization.

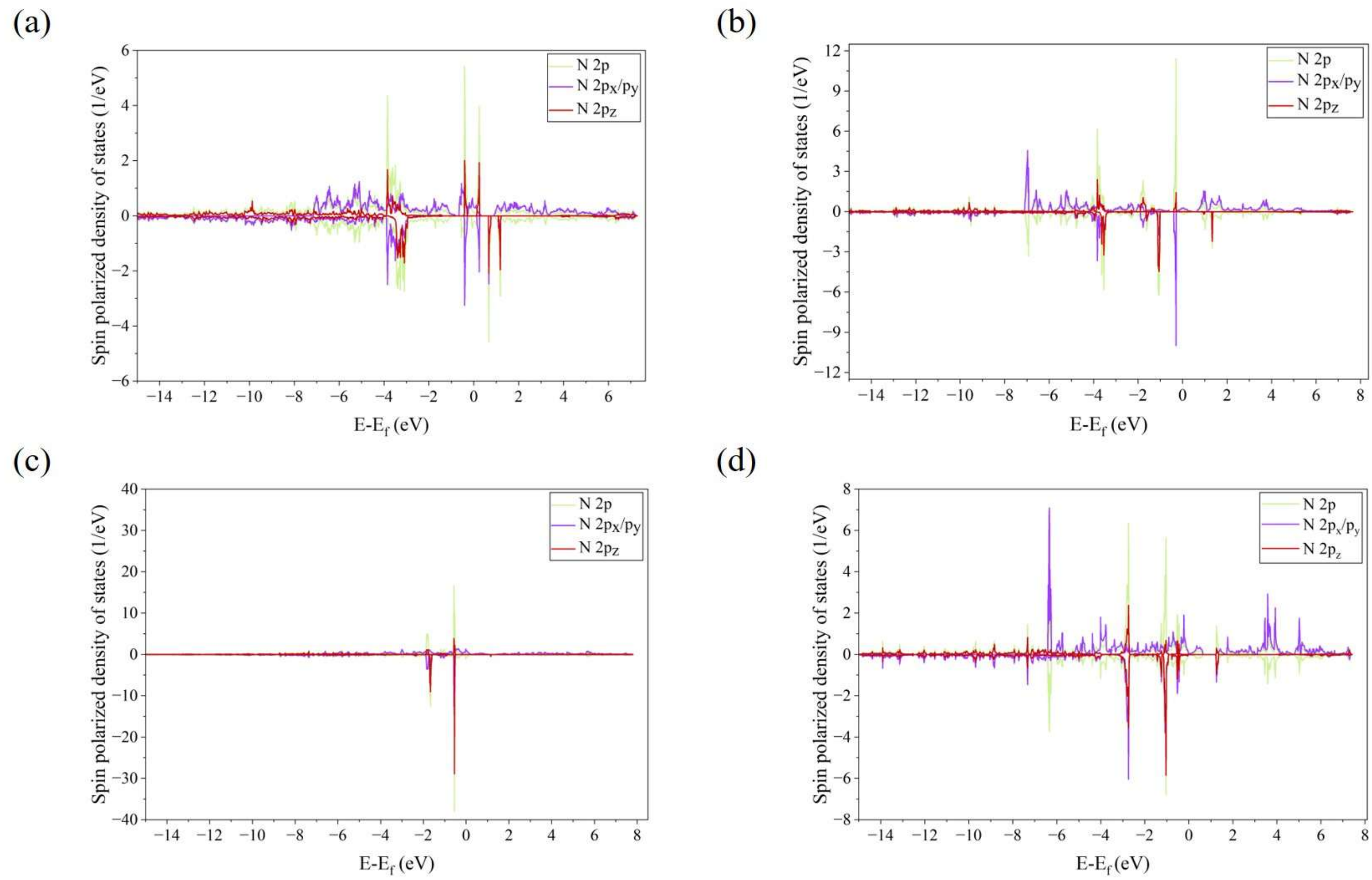

**Figure 8.** N-atom-resolved SPDOS showing N 2p, N (2p_x/2p_y), and N (2p_z) contributions for the (a) pyridinic, (b) pyridazinic, (c) pyrrolic, and (d) pyrazolic configurations.

The magnetic contrast among the four systems originates from topology-controlled occupation of vacancy-edge $p_z$ states. In the pyridinic, pyridazinic, and pyrrolic configurations, the asymmetric edge bonding and uneven charge redistribution leave uncompensated localized $p_z$ electrons on neighboring carbon atoms, generating finite spin-polarized moments. Nitrogen therefore acts not as the sole magnetic center, but as a spin-modulating dopant that polarizes the surrounding defect-edge carbon network. In the pyrazolic configuration, however, the more balanced N-pair coordination redistributes charge in a way that pairs the localized valence electrons and compensates the spin-up and spin-down contributions, resulting in zero net magnetization.

## Conclusion

This work systematically shows that the structural, electronic, and magnetic characteristics of a nitrogen-doped defective graphene are not just controlled by the presence of nitrogen but by the exact topology of the peripheral nitrogen atoms. Four configurations of the N atoms- pyridinic, pyridazinic, pyrrolic, and pyrazolic were compared in a consistent computational protocol using spin-polarized first-principles calculations. The peripheral N-motif each produces a unique reconstruction pattern around the central void, as revealed by the structural analysis. The calculated defect formation energies show the thermodynamic stability sequence as: pyridinic > pyrrolic > pyridazinic > pyrazolic. The band structures and calculated density of states indicate that pyridinic, pyridazinic and pyrrolic configurations remain metallic or near-metallic. In contrast, the pyrazolic configuration opens a PBE-level band gap of 0.377 eV. Spin-density and local magnetic moment analyses have shown finite magnetic moments for pyridinic, pyridazinic, and pyrrolic configurations. The pyrazolic configuration, on the other hand, shows total spin compensation. These results demonstrate that the peripheral parts of the N topology can provide a defect-chemical switch to select between magnetic metallic/near-metallic and nonmagnetic semiconducting graphene states. The results offer a basic design principle to engineer nitrogen-doped graphene with desired electronic and spin-dependent properties for future use in nanoelectronic, sensing, and spintronic devices. Future studies may further enhance this framework by calculating the ferromagnetic and antiferromagnetic ordering energies, exchange coupling, estimates of the Curie temperature, hybrid-functional band gaps, and finite-size effects. Such additional analyses would help connect the local spin-polarized defect states identified here with experimentally accessible long-range magnetic and transport behavior.